\documentclass[twocolumn,showpacs,preprintnumbers,amsmath,amssymb, a4]{revtex4}
\usepackage[dvipdfmx]{graphicx}
\usepackage{dcolumn}
\usepackage{amsmath}
\usepackage{multirow}
\usepackage{color}
\usepackage{bm}

\newcommand{\TC}{$T_{\rm C}$}

\begin{document}

\title{A regression-based feature selection study of the Curie temperature of transition-metal rare-earth compounds: prediction and understanding}

\author{
Hieu Chi Dam$^{1,2,3}$, Viet Cuong Nguyen$^4$, Tien Lam Pham$^{1,5}$, Anh Tuan Nguyen$^6$,  \\
Hiori Kino$^{2,5}$,  Kiyoyuki Terakura$^{1,2}$, and Takashi Miyake$^{2,5,7}$ 
}
\affiliation{
$^{1}$Japan Advanced Institute of Science and Technology, 1-1 Asahidai, Nomi,Ishikawa 923-1292, Japan\\
$^{2}$CMI$^2$, MaDIS, NIMS, Tsukuba, Ibaraki 305-0047, Japan\\
$^{3}$JST, PRESTO, 4-1-8 Honcho, Kawaguchi, Saitama, 332-0012, Japan\\
$^{4}$HPC Systems Inc., Japan\\
$^{5}$ESICMM, NIMS, Tsukuba 305-0047, Japan \\
$^{6}$Hanoi Metropolitan University, 98 Duong Quang Ham, Cau Giay, Hanoi, Vietnam \\
$^{7}$CD-FMat, AIST, Tsukuba, Ibaraki 305-8568, Japan
}

\date{\today}

\begin{abstract}
The Curie temperature ({\TC}) of binary alloy compounds consisting of 3$d$ transition-metal and 4$f$ rare-earth elements is analyzed by a machine learning technique. We first demonstrate that nonlinear regression can accurately reproduce {\TC} of the compounds. The prediction accuracy for {\TC} is maximized when five to ten descriptors are selected, with the rare-earth concentration being the most relevant. We then discuss an attempt to utilize a regression-based model selection technique to learn the relation between the descriptors and the actuation mechanism of the corresponding physical phenomenon, i.e.,{\TC} in the present case. 
\end{abstract}

 
\maketitle
The development of strong permanent magnets is an urgent technological issue as well as a fundamental challenge in materials science. Most strong permanent magnets are rare-earth magnets that mainly consist of transition-metal and rare-earth elements. To date, various rare-earth transition-metal compounds have been synthesized. The strongest magnetic compound is Nd$_{2}$Fe$_{14}$B developed by Sagawa et al., which is the main phase of neodymium magnets \cite{sagawa84}. Although Sagawa's intention of raising {\TC} by adding B was successful ({\TC} = 335 K (Nd$_2$Fe$_{17}$) vs. 586 K (Nd$_2$Fe$_{14}$B)),  {\TC} of Nd$_2$Fe$_{14}$B is still much lower than those of Sm$_{2}$Co$_{17}$ (1,193~K) and $\alpha$-Fe (1,043~K). Because of this, dysprosium is added to ensure a sufficient coercivity at high temperatures for technological usage. 

It is highly desirable to establish a technique that enables an accurate description of {\TC} and to clarify the controlling parameters (descriptors) that influence {\TC} for the development of a new strong magnet. 
From a theoretical point of view, an accurate description of {\TC} is a demanding task. The spin magnetic moment at each atomic site and the magnetic exchange coupling between them can be evaluated within the framework of density functional theory \cite{calculation_magnetic_moment_Liu2011841}. The Curie temperature can be obtained by solving a derived classical spin model. 
It is known that this approach works reasonably well for systems where the longitudinal spin fluctuation can be neglected. In the case of rare-earth compounds, however, the 4$f$ electrons bring further complications. Both the electron correlation and spin--orbit interaction are strong in these compounds; hence, an advanced theoretical treatment is needed for a reliable description of the magnetic properties. 

The recent rapid development of data-driven approaches in materials research offers another possibility \cite{data_mining_materials_science_PRB2012, identifying_zeolite_framework_JPC_C2012, find_missing_ternary_oxide_chem_mater_2012, find_DFT_PRL_2012, Materials_cartography_chem_mater_2015}. Nowadays, we are able to obtain {\TC} data for many compounds from databases or the literature. Machine learning may be utilized to predict {\TC} of a new compound from the existing data. This approach is particularly suited to materials exploration because of its high efficiency. In the present work, we demonstrate how this idea is achieved by employing the Gaussian kernel ridge regression (GKR) technique for machine learning. We focus on {\TC} of transition-metal rare-earth bimetals. We present a comparison between the value of {\TC} predicted by GKR and the observed data and analyze the selection of descriptors for obtaining a model for predicting {\TC} with a high prediction accuracy. In this way, we obtain information about the important descriptors for the {\TC} prediction. We may also expect that insights into the actuating mechanisms of physical phenomena can be obtained through the process mentioned above. However, ``{\it prediction}'' and ``{\it understanding}'' are not necessarily achieved simultaneously.  This is because the ``{\it correlation relation}'' obtained by the machine learning prediction is generally different from the ``{\it causality relation}'' required for the understanding of the actuating mechanism.
In the following, we first discuss how a prediction with a high accuracy can be achieved and then present our attempts to obtain insights into the actuating mechanism.\\

\clearpage
\begin{center}
{\it Preparation for the data analysis}
\end{center}

We collected the experimental data of 108 binary compounds consisting of transition metals and rare-earth metals from the AtomWork database of NIMS \cite{paulingfile, atomwork}, including the crystal structure of the compounds and their observed {\TC}. Our task is to learn a model for predicting {\TC} of a new compound from the training data of known compounds. For this purpose, one of the most important steps is the choice of an appropriate data representation that reflects knowledge of the application domain, i.e., a model of the underlying physics. 

To represent the structural and physical properties of each binary compound, we use a combination of 28 descriptors. We divide all 28 descriptors into three categories, as summarized in Table \ref{tab.Descriptors}. 

\begin{table}
\centering
\caption{Descriptors representing the structural and physical properties of binary compounds. (Details can be found in Section A in the Supplemental Materials \cite{supplemental}).}
\begin{tabular}{p{3.1cm}|p{5.1cm}}
\hline\hline
\centering
Category &Descriptors \\
\hline
Atomic properties of transition metals  ($\bm{\bar{T}}$) & $Z_T$, $r_T$, $r^{cv}_{T}$, $IP_{T}$, $\chi_{T}$, $S_{3d}$, $L_{3d}$, $J_{3d}$\\
\hline
Atomic properties of rare-earth metals  ($\bm{\bar{R}}$)  & $Z_R$, $r_{R}$, $r^{cv}_{R}$, $IP_{R}$, $\chi_{R}$, $S_{4f}$, $L_{4f}$, $J_{4f}$, $g_J$, $J_{4f}g_{J}$, $J_{4f}(1-g_J)$\\
\hline
Structural information ($\bm{\bar{S}}$)  &$C_T$, $C_R$, $d_{T-T}$, $d_{T-R}$, $d_{R-R}$, $N_{T-T}$, $N_{T-R}$, $N_{R-R}$, $N_{R-T}$\\
\hline\hline
\end{tabular}
\label{tab.Descriptors} 
\end{table}

The first and second categories pertain to the descriptors describing the atomic properties of the transition-metal elements ($\bm{\bar{T}}$ descriptors) and rare-earth elements ($\bm{\bar{R}}$ descriptors), respectively. The descriptors related to the magnetic properties are included. 
It has been well established that information related to the crystal structure is very valuable in relation to understanding the physics of binary compounds with transition metals and rare-earth metals. Therefore, we design the third category with structural descriptors ($\bm{\bar{S}}$ descriptors) whose values are calculated from the crystal structures of the compounds in the literature. 
Note that the concentration of the transition metal ($C_T$) and that of the rare-earth metal ($C_R$) are both used. If we use the atomic percent for the concentration, the two quantities are not independent. Here, we measure the concentration in units of (atoms/{\AA}$^3$), which is more informative than the atomic percent because the former contains information about the constituent atomic size. Then, ($C_R$) and ($C_T$) are not totally dependent.

\begin{center}
{\it Prediction}
\end{center}

To learn a function for predicting the values of {\TC} of compounds from the data represented by using vectors of the descriptors, we apply the GKR technique \cite{ML}, which has recently been applied successfully to many materials science issues \cite{Rupp_tutorial, Botu, Pilania}. For GKR, the predicted property $f(\bm{x})$ at the point $\bm{x}$ is expressed as the weighted sum of Gaussians:
\begin{equation}
f(\bm{x}) = \Sigma_{i=1}^N c_i \exp (\frac{-\|\bm{x}_i-\bm{x}\|^2_2}{2\sigma^2}),  
\label{eq.fx}
\end{equation}
where $N$ is the number of training data points, $\sigma^2$ is a parameter corresponding to the variance of the Gaussian kernel function, and $\|\bm{x}_i-\bm{x}\|^2_2 = \Sigma_{\alpha=0}^{N_D}(x_i^{\alpha}-x^{\alpha})^2$ is the squared $L^2$ norm of the difference between the two $N_D$-dimensional descriptor vectors $\bm{x_i}$ and $\bm{x}$. The coefficients $c_i$ are determined by minimizing
\begin{equation}
\Sigma_{i=1}^N [f(\bm{x_i}) - y_i]^2 + \lambda\Sigma_{i=1}^{N} ||c_i||_2^2,  
\label{eq.EL}
\end{equation}
where $y_i$ is the observed data value.
The regularization parameters $\lambda$ and $\sigma$ are chosen with the help of cross-validation, i.e., by excluding some of the materials as a test set during the training process and measuring the coefficient of determination $R^2$ defined by
\begin{equation}
R^2=1-\frac
{\Sigma_{j=1}^{N_{\mathrm{test}}} [y_j - f(\bm{x_j})]^2}
{\Sigma_{j=1}^{N_{\mathrm{test}}} [y_j-\bar{y}]^2},
\end{equation}
where $N_{\mathrm{test}}$ is the number of data points, and $\bar{y}$ is the average of the test set
for testing how the predicted values for the excluded materials agree with the actually observed values. In this study, we use $R^2$ as a measure of the prediction accuracy. To obtain a good estimate of the prediction accuracy, we carried out GKR with 100 times 10-fold cross-validation using the collected data, of which {\TC} is the target quantity.

To find the most appropriate set of descriptors for the prediction of {\TC}, we train the GKR models for all combinations of the descriptors. With each combination, we search for the regularization parameters $\lambda$ and $\sigma$ to maximize the prediction accuracy. Then, we obtain new data for all possible sets of descriptors and the corresponding best prediction accuracy that the GKR model can achieve. By analyzing these new data, we aim 
(1) to find the set of descriptors that yields the best prediction accuracy for the prediction of {\TC} and
(2) to find the descriptors with strong relevance for the prediction of {\TC}.

We define the prediction ability $PA(\bm{S})$ 
of descriptors by the maximum prediction accuracy that the GKR model can achieve by using the descriptors in a subset $\bm{s}$ of a set $\bm{S}$ of descriptors as follows:

\begin{equation}
PA(\bm{S}) = \max_{\forall \bm{s} \subset \bm{S}} R^2_s, 
\label{eq.PAS}
\end{equation}
where $R^2_s$ is the value of the coefficient of determination $R^2$ achieved by GKR using a descriptor set $s$.



\begin{figure}
\centering
\includegraphics[width=0.42\textwidth]{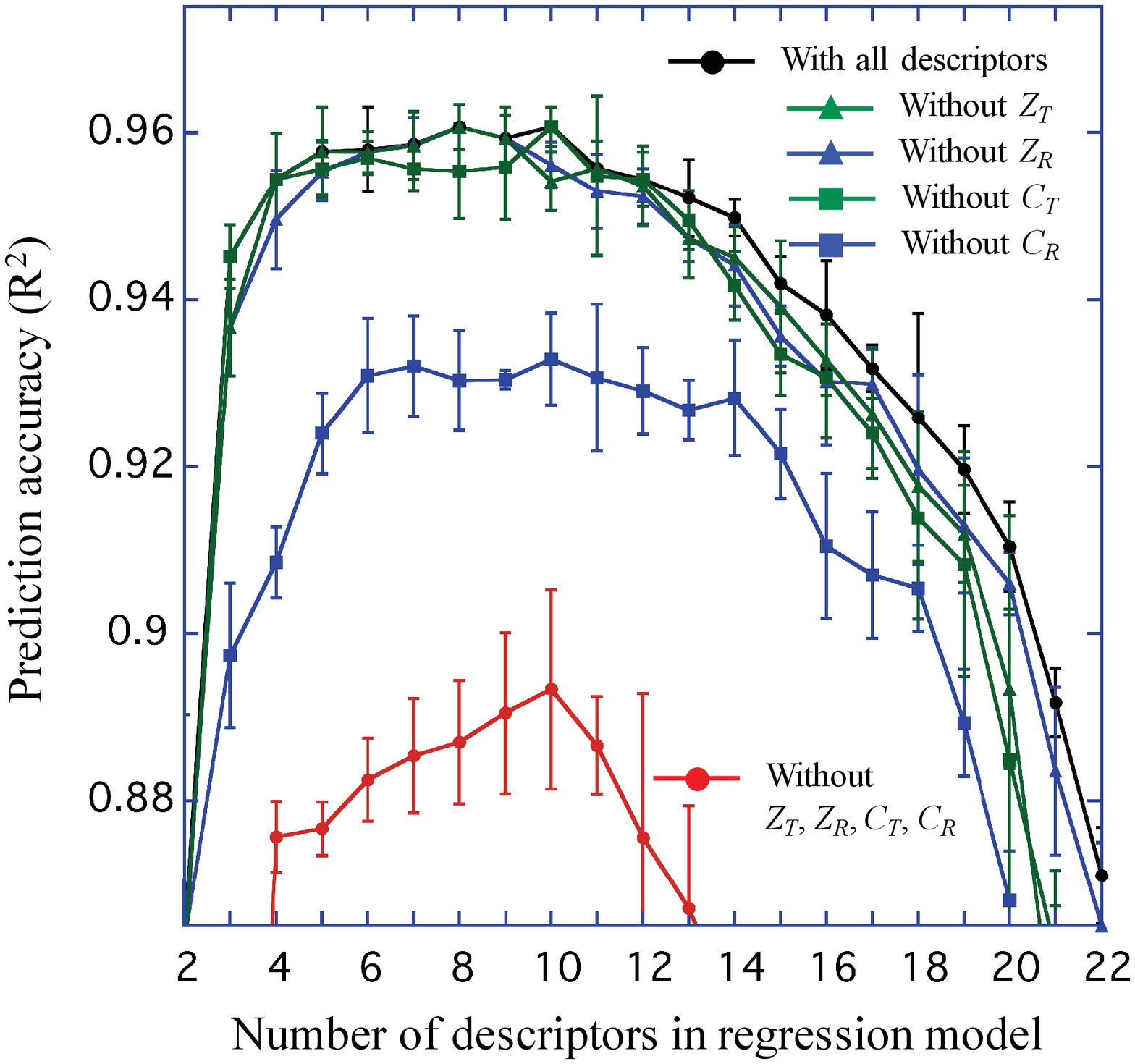}
\caption{Dependence of the best prediction accuracy on the number of descriptors in the Gaussian kernel regression model. The black solid circles and black line represent the results for the full set of descriptors. The red solid circles and red line represent the results when none of the four descriptors $Z_T$, $Z_R$, $C_T$, and $C_R$ are recruited. The green solid triangles, blues solid triangles, green solid squares, blue solid squares, and the corresponding lines represent the results when the $Z_T$, $Z_R$, $C_T$, or $C_R$ descriptor is removed, respectively.}

\label{fig.numD}
\end{figure}

\begin{figure}
\centering
\includegraphics[width=0.42\textwidth]{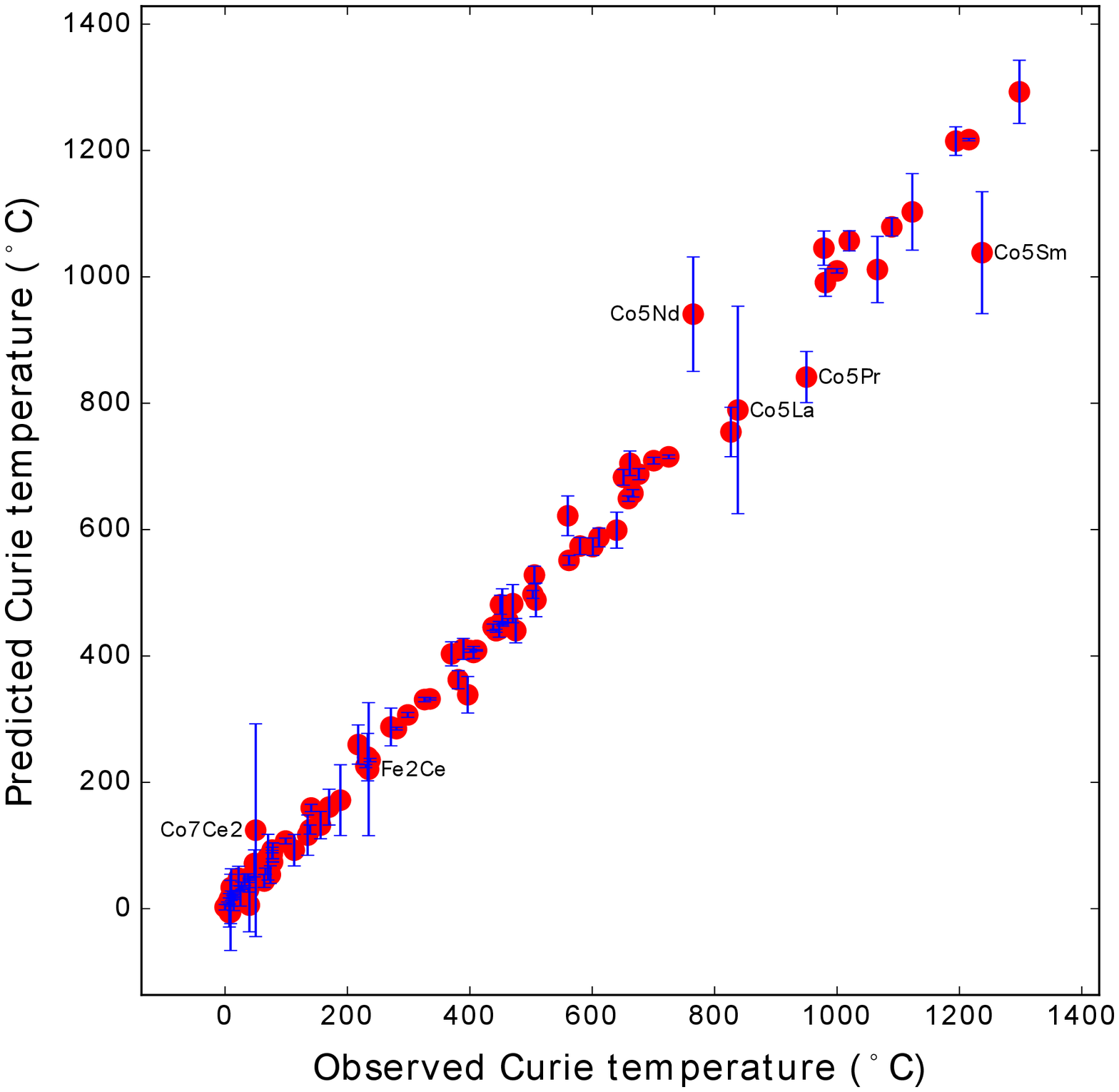} 
\caption{Observed and predicted (by nonlinear regression) Curie temperatures for 108 bimetal alloy compounds. The black solid circles represent the results of the best Gaussian kernel ridge regression using eight descriptors. The blue lines represent the error margins of the prediction estimated by cross-validation.}
\label{fig.Tc}
\end{figure}

Once a large amount of new data (data of the GKR models) is obtained ($2^{N_D}-1$ with $N_D=28$), we are ready to analyze these new data to find the most appropriate model for the prediction of {\TC}. We first examine how {\TC} can be predicted by using the designed descriptors of the compounds. Figure \ref{fig.numD} shows the dependence of the best prediction accuracy on the number of descriptors recruited in the GKR model. The results indicate that it is possible to accurately predict {\TC} of rare-earth transition-metal bimetal alloys by using GKR with the designed descriptors. The prediction accuracy reaches a maximum with five to ten descriptors (see Table \ref{tab.Bestdescriptors} in the Supplemental Materials \cite{supplemental}). Note that for each fixed number of descriptors recruited in GKR, a couple of different sets of descriptors also attain a high prediction accuracy that is almost comparable to the best one because the descriptors are not independent. These sets are listed in Table \ref{tab.Bestdescriptors} in the Supplemental Materials. By using a set of eight descriptors ($C_R$, $C_T$, $Z_R$, $Z_T$, $IP_{T}$, $S_{3d}$, $J_{3d}$, and $L_{3d}$), we can obtain an excellent prediction accuracy (as seen in Fig.~\ref{fig.Tc}) with $R^2$ and the mean absolute error (MAE) being approximately 0.96 and 41~K, respectively. It is clear that the prediction accuracy gradually decreases when the number of recruited descriptors increases. This result originates from the fact that the overuse of many weakly relevant descriptors weakens the correlation between the similarity of the compounds, which is measured using the Gaussian kernel of the descriptors and the differences in their values of {\TC}.

\begin{figure}
\centering
\includegraphics[width=0.40\textwidth]{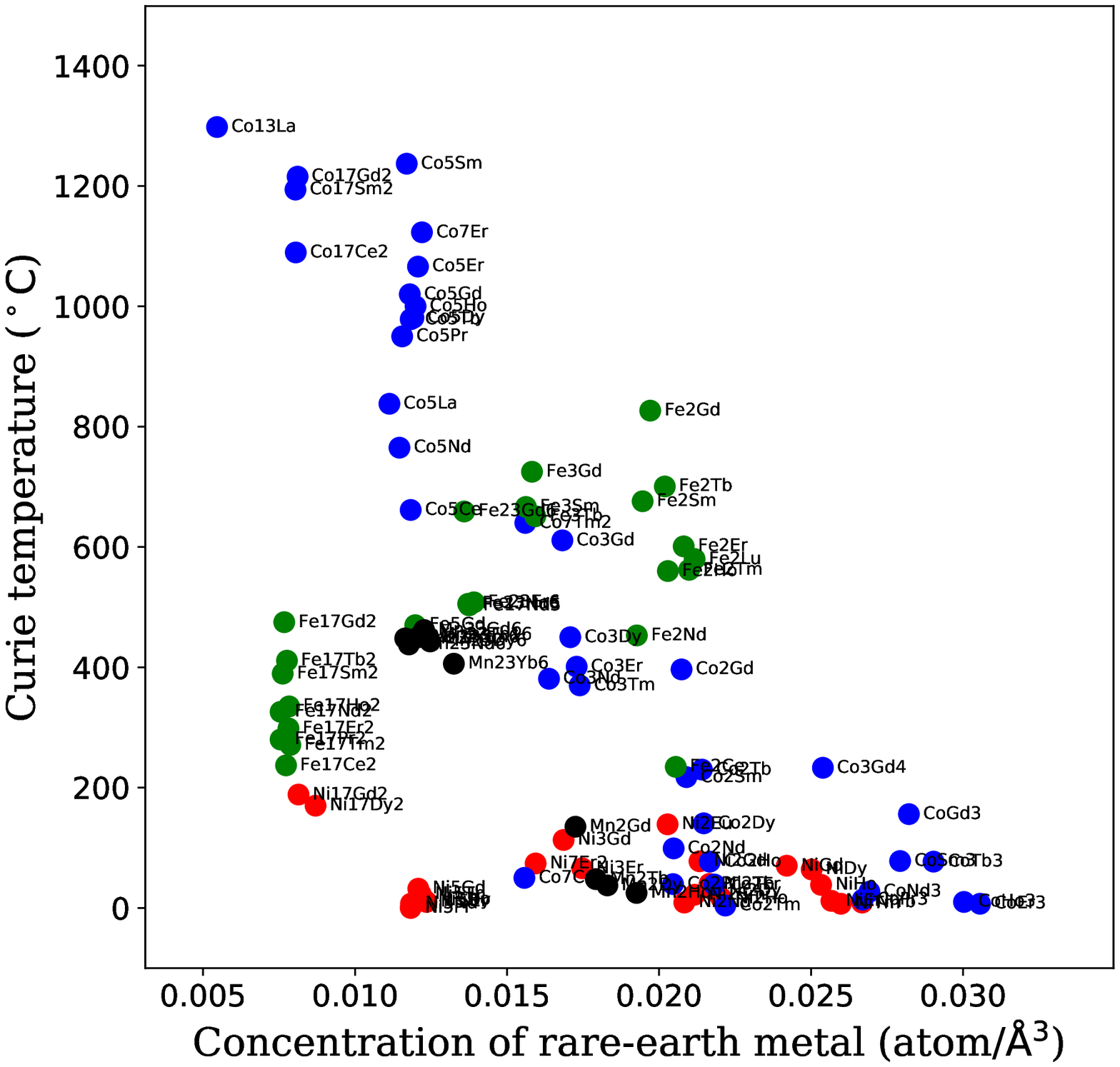}
\caption{Dependence of the Curie temperature ({\TC}) on the concentration of the rare-earth metal ($C_R$) in binary alloy compounds. The black, green, blue, and red solid circles represent the alloys of Mn, Fe, Co, and Ni, respectively.
}
\label{fig.CR}
\end{figure}

Next, we evaluate the relevance of each descriptor for the prediction of {\TC}. We compare $PA(\bm{S})$ of the full set of descriptors $\bm{S}$ ($N_S=28$) and $PA(\bm{S}-\{d_i\})$ for  of the descriptors $d_i$.  We call this a leave-one-out test.
We find that most of the descriptors are weakly relevant (see the Supplemental Materials for the definitions of strong and weak relevance and irrelevance \cite{supplemental}), and the prediction accuracy does not significantly change, except in one case for $C_R$. Figure~\ref{fig.numD} shows some examples of the leave-one-out test when $\{d_i\}$ is either $C_R$, $Z_R$, $C_T$, or $Z_T$. It is clearly seen that the absence of $C_R$ in the GKR model results in a dramatic decrease in the accuracy: $PA(\bm{S}) > PA(\bm{S}-\{C_R\})$; therefore, $C_R$ is surely assigned as a strongly relevant descriptor in terms of the prediction of {\TC}. 
On the other hand, for the other three descriptors $Z_R$, $C_T$, and $Z_T$, only a marginal reduction in the average of the prediction accuracy can be observed, particularly for the number of descriptors ranging from five to ten. We can conclude that the information embedded in each of $Z_R$, $C_T$, and $Z_T$ can be compensated by other descriptors, whereas the information embedded in $C_R$ cannot be compensated by other descriptors.
The essentiality of $C_R$ can be confirmed easily in Fig.~\ref{fig.CR}, where the upper limit of {\TC} linearly depends on $C_R$. Note, however, that the $C_R$ dependence of {\TC} is even qualitatively different among transition-metal counterparts.  For Mn and Co as transition metals, {\TC} tends to decrease with $C_R$; however, it tends to increase for Fe. For Ni, {\TC} is rather insensitive to $C_R$. It is important to note that the GKR model can reproduce the situation quite well.


\begin{center}
{\it Understanding}
\end{center}

Since the embedded information in a weakly relevant descriptor can be recovered by other descriptors, it is important to compare the degrees of relevance of descriptors within each category for the prediction of {\TC} to obtain an insight into the mechanism determining {\TC}. This can be achieved with an add-one-in test, in which we first remove all of the descriptors belonging to one of the three categories of descriptors and evaluate $R^2$ using only the descriptors of the remaining two categories.  Then, we estimate $R^2$ by recovering each of the removed descriptors. This test enables us to know which descriptor is the most relevant in each category. At the same time, we choose the minimum set of descriptors among the removed descriptors to recover nearly the same value of the highest $R^2$. The results of these tests are described below for each of the three categories of descriptor sets in Table \ref{tab.Descriptors}.

1) Structural descriptors ($\bm{\bar{S}}$ descriptors): 
  We first remove all of the descriptors describing the crystal structures ($\bm{\bar{S}}$ descriptors). Then, as the rest of descriptors cannot distinguish different materials properly, $R^2$ becomes 0.0. The results of the add-one-in test in the order of high $R^2$ are presented in the first two columns of Table \ref{tab.R2scores} for each of the $\bm{\bar{S}}$ descriptors. These $R^2$ values were obtained with two different methods using the descriptors in the other two categories: 1) the $R^2$ values obtained for the optimal choice of the $\bm{\bar{R}}$ and $\bm{\bar{T}}$ descriptors are shown in parentheses in Table \ref{tab.R2scores}, and 2) those obtained by excluding $Z_R$, $\chi_R$ (electronegativity), and $Z_T$ are shown without parentheses. The reasoning behind the second method will be discussed below. $R^2$ only for $C_R$ suggests that the $\bm{\bar{S}}$ descriptors can be well represented solely by $C_R$. This is consistent with the above assignment of $C_R$ as a descriptor with strong relevance. Therefore, in the add-one-in tests for the $\bm{\bar{R}}$ and $\bm{\bar{T}}$ descriptors, we use only $C_R$ for the $\bm{\bar{S}}$ descriptors.

2) Atomic descriptors for rare-earth elements ($\bm{\bar{R}}$ descriptors):
 The third and fourth columns of Table \ref{tab.R2scores} present the results of the add-one-in test for the $\bm{\bar{R}}$ descriptors. With only $Z_R$, $R^2$ reaches 0.956 (0.957); any possible combinations of $\bm{\bar{R}}$ descriptors without $Z_R$ can only give an $R^2$ of 0.947 (0.950). Therefore, $Z_R$ can represent the entire set of $\bm{\bar{R}}$ descriptors for prediction. In the present problem, as $Z_R$ covers only the $4f$ series and {\TC} mostly varies in systematic ways across the rare-earth-element series in the experimental data, rare-earth elements can be suitably identified with $Z_R$. If $Z_R$ covers more of the $f$ series with different principal quantum numbers, $Z_R$ will not be a suitable descriptor for measuring the similarity among elements. Moreover, according to our domain knowledge, $Z_R$ itself will not directly appear in the physical model for {\TC}. From a causal analysis, $Z_R$ gives information about the electronic configuration based on which magnetic properties are evaluated. Table \ref{tab.R2scores} also indicates that $\chi_R$ serves as a similarly relevant $\bm{\bar{R}}$ descriptor. Note, however, that this is simply due to its nearly linear relation with $Z_R$. On the basis of such a consideration, we remove $Z_R$ and $\chi_R$ from the $\bm{\bar{R}}$ descriptors for understanding and look for alternative physical quantities as relevant $\bm{\bar{R}}$ descriptors. This is the reason of why we adopt the second method for choosing descriptors described above. We see from Table \ref{tab.R2scores} that $J_{4f}(g_J-1)$ gives nearly the same $R^2$ as that for $Z_R$ and that it is more relevant than $S_{4f}$ and $J_{4f}$. Moreover, the last line of the third and fourth columns of Table \ref{tab.R2scores} suggests that $J_{4f}(g_J-1)$ can only represent the $\bm{\bar{R}}$ descriptors without $Z_R$.
  In the $4f$ state of the rare-earth element, the spin--orbit interaction is strong, the total angular momentum $J_{4f}$ rather than each of $S_{4f}$ and $L_{4f}$ is conserved, and $J_{4f}(g_J-1)$ is the projection of $S_{4f}$ onto $J_{4f}$. The Curie temperature {\TC} will be controlled by the effective $4f$--$3d$ exchange interaction through the RE-$5d$ states in which $J_{4f}(g_J-1)$ plays a role. With this physical picture, the above result of the add-one-in test is physically meaningful.

3) Atomic descriptors for transition-metal elements ($\bm{\bar{T}}$ descriptors):
 The results of the add-one-in test for the $\bm{\bar{T}}$ descriptors are summarized in the last two columns of Table \ref{tab.R2scores}. Similar to the $\bm{\bar{R}}$ descriptor case, although the atomic number $Z_T$ is the best single choice of the $\bm{\bar{T}}$ descriptors for prediction in the present problem, we search for a more physically meaningful choice of $\bm{\bar{T}}$ descriptors. The results in Table \ref{tab.R2scores} clearly show that $S_{3d}$ is much more relevant than $J_{3d}$ for $3d$ transition-metal elements. This is again physically meaningful because of the weak spin--orbit interaction and significant $3d$ band width. From Table \ref{tab.R2scores}, we can also see that the combination of $S_{3d}$ and $IP_T$ gives an even higher $R^2$ than that for $Z_T$. This is also physically meaningful because {\TC} can be controlled not only by $S_{3d}$ but also by the $3d$--$3d$ exchange interaction, which may be affected by $IP_T$ (related to the $3d$ energy level). 

\begin{table*}
\centering
\caption{Dependence of the prediction accuracy of the GKR model on the descriptors in the add-one-in test.}
\label{my-label}
\begin{tabular}{cccccc}
\hline 
\multicolumn{2}{|c|}{$\bm{\bar{S}}$ descriptors}                              & \multicolumn{2}{c|}{$\bm{\bar{R}}$ descriptors}                              & \multicolumn{2}{c|}{$\bm{\bar{T}}$ descriptors}                                  \\ 
\multicolumn{2}{|c|}{}                                                        & \multicolumn{2}{c|}{only $C_R$ for $\bm{\bar{S}}$ descriptor}                & \multicolumn{2}{c|}{only $C_R$ for $\bm{\bar{S}}$ descriptor}                    \\ \hline
\multicolumn{1}{|c|}{descriptors in use} & \multicolumn{1}{c|}{$R^2$}         & \multicolumn{1}{c|}{descriptors in use} & \multicolumn{1}{c|}{$R^2$}         & \multicolumn{1}{c|}{descriptors in use}             & \multicolumn{1}{c|}{$R^2$} \\ \hline
\multicolumn{1}{|c|}{none}               & \multicolumn{1}{c|}{0.0}           & \multicolumn{1}{c|}{none}               & \multicolumn{1}{c|}{0.0}           & \multicolumn{1}{c|}{none}                           & \multicolumn{1}{c|}{0.265} \\ 
\multicolumn{1}{|c|}{$C_R$}              & \multicolumn{1}{c|}{0.940 (0.957)} & \multicolumn{1}{c|}{$Z_R$}              & \multicolumn{1}{c|}{0.956 (0.957)} & \multicolumn{1}{c|}{$Z_T$}                          & \multicolumn{1}{c|}{0.945} \\ \multicolumn{1}{|c|}{$C_T$}              & \multicolumn{1}{c|}{0.909 (0.911)} & \multicolumn{1}{c|}{$\chi_R$}           & \multicolumn{1}{c|}{0.945 (0.950)} & \multicolumn{1}{c|}{$S_{3d}$}                       & \multicolumn{1}{c|}{0.937} \\ 
\multicolumn{1}{|c|}{$d_{R-R}$}          & \multicolumn{1}{c|}{0.876 (0.882)} & \multicolumn{1}{c|}{$J_{4f}(g_J-1)$}    & \multicolumn{1}{c|}{0.940 (0.950)} & \multicolumn{1}{c|}{$IP_T$}                         & \multicolumn{1}{c|}{0.894} \\ 
\multicolumn{1}{|c|}{$d_{T-T}$}          & \multicolumn{1}{c|}{0.621 (0.632)} & \multicolumn{1}{c|}{$S_{4f}$}           & \multicolumn{1}{c|}{0.919 (0.920)} & \multicolumn{1}{c|}{$\chi_T$}                       & \multicolumn{1}{c|}{0.883} \\ 
\multicolumn{1}{|c|}{$d_{T-R}$}          & \multicolumn{1}{c|}{0.518 (0.518)} & \multicolumn{1}{c|}{$g_J$}              & \multicolumn{1}{c|}{0.909 (0.911)} & \multicolumn{1}{c|}{$J_{3d}$}                       & \multicolumn{1}{c|}{0.702} \\ 
\multicolumn{1}{|c|}{$C_R, C_T$}         & \multicolumn{1}{c|}{0.934 (0.960)} & \multicolumn{1}{c|}{$J_{4f}g_J$}        & \multicolumn{1}{c|}{0.905 (0.907)} & \multicolumn{1}{c|}{$L_{3d}$}                       & \multicolumn{1}{c|}{0.426} \\ 
\multicolumn{1}{|c|}{}                   & \multicolumn{1}{c|}{}              & \multicolumn{1}{c|}{$IP_R$}             & \multicolumn{1}{c|}{0.892 (0.894)} & \multicolumn{1}{c|}{$IP_T$, $S_{3d}$}               & \multicolumn{1}{c|}{0.950} \\ 
\multicolumn{1}{|c|}{}                   & \multicolumn{1}{c|}{}              & \multicolumn{1}{c|}{$J_{4f}$}           & \multicolumn{1}{c|}{0.871 (0.870)} & \multicolumn{1}{c|}{$IP_T$, $S_{3d}$, $J_{3d}$}     & \multicolumn{1}{c|}{0.956} \\ 
\multicolumn{1}{|c|}{}                   & \multicolumn{1}{c|}{}              & \multicolumn{1}{c|}{$L_{4f}$}           & \multicolumn{1}{c|}{0.0 (0.0)}     & \multicolumn{1}{c|}{any $\bm{\bar{T}}$ descriptor} & \multicolumn{1}{c|}{0.957} \\ 
\multicolumn{1}{|c|}{}                   & \multicolumn{1}{c|}{}              & \multicolumn{1}{c|}{any without $Z_R$}  & \multicolumn{1}{c|}{0.947 (0.950)} & \multicolumn{1}{c|}{}                               & \multicolumn{1}{c|}{}      \\ \hline
\multicolumn{1}{l}{}                     & \multicolumn{1}{l}{}               & \multicolumn{1}{l}{}                    & \multicolumn{1}{l}{}               & \multicolumn{1}{l}{}                                & \multicolumn{1}{l}{}       \\
\multicolumn{1}{l}{}                     & \multicolumn{1}{l}{}               & \multicolumn{1}{l}{}                    & \multicolumn{1}{l}{}               & \multicolumn{1}{l}{}                                & \multicolumn{1}{l}{}       \\
\multicolumn{1}{l}{}                     & \multicolumn{1}{l}{}               & \multicolumn{1}{l}{}                    & \multicolumn{1}{l}{}               & \multicolumn{1}{l}{}                                & \multicolumn{1}{l}{}      
\end{tabular}
\label{tab.R2scores}
\end{table*}

Summarizing the above results, we select the following four descriptors compatible with the physical model of {\TC}: $C_R$, $J_{4f}(g_J-1)$, $S_{3d}$, and $IP_T$. This set of descriptors leads to $R^2=0.940$, which is sufficiently high.   

\begin{center}
{\it Conclusion}
\end{center}

In this study, we analyze the Curie temperatures of binary alloys consisting of 3$d$ transition-metal and 4$f$ rare-earth elements by a machine learning technique. Nonlinear regression with a Gaussian kernel accurately reproduces the values {\TC} of 108 compounds. A regression-based model selection technique is utilized for learning the relation between the descriptors and {\TC}. The prediction accuracy of {\TC} is maximized when eight descriptors are selected, with the rare-earth concentration being the most relevant. We demonstrate how the regression-based model selection technique can be utilized for mining the scientific connection between the descriptor and the actuation mechanisms of a physical phenomenon.
\begin{acknowledgments}
This work was partly supported by PRESTO and by the ``Materials Research by Information Integration'' Initiative (MI$^2$I) project of the Support Program for Starting Up Innovation Hub, both from the Japan Science and Technology Agency (JST), Japan; by the Elements Strategy Initiative Project under the auspices of MEXT; and also by 
MEXT as a social and scientific priority issue (Creation of New Functional Devices and High-Performance Materials to 
Support Next-Generation Industries; CDMSI) to be tackled by using a post-K computer.
\end{acknowledgments}


\clearpage

\section{Supplemental Materials}

\subsection{Data collection and data representation} 
We collected the experimental data of 108 binary compounds consisting of transition metals and rare-earth metals from the Atomwork database of NIMS \cite{paulingfile, atomwork}, including the crystal structure of the compounds and their observed {\TC}. To represent the structural and physical properties of each binary compound, we use a combination of 28 descriptors. We divide all 28 descriptors into three categories. 

The first category pertains to the descriptors describing the atomic properties of the transition-metal constituent, including the (1) atomic number ($Z_T$), (2) atomic radius ($r_T$), (3) covalent radius ($r^{cv}_{T}$), (4) ionization potential ($IP_{T}$), (5) electronegativity ($\chi_{T}$), (6) spin angular moment ($S_{3d}$), (7) orbital angular moment ($L_{3d}$), and (8) total angular moment ($J_{3d}$) of the 3$d$ electrons. The selection of these descriptors originates from the physical consideration that the intrinsic electronic and magnetic properties will determine the 3$d$ orbital splitting at transition-metal sites. 

In the same manner, we design the second category pertaining to the descriptors for describing the properties of the rare-earth metal constituent, including the (9) atomic number ($Z_R$), (10) atomic radius ($r_{R}$), (11) covalent radius ($r^{cv}_{R}$), (12) ionization potential ($IP_{R}$), (13) electronegativity ($\chi_{R}$), (14) spin angular moment ($S_{4f}$), (15) orbital angular moment ($L_{4f}$), and (16) total angular moment ($J_{4f}$) of the 4$f$ electrons. To capture the effect of the $4f$ electrons better, we add three additional descriptors for describing the properties of the constituent rare-earth metal ions, including (17) the Land$\acute{e}$ factor ($g_J$), (18) the projection of the total magnetic moment onto the total angular moment ($J_{4f}g_{J}$), and (19) the projection of the spin magnetic moment onto the total angular moment ($J_{4f}(1-g_J)$) of the 4$f$ electrons. The selection of these features originates from the physical consideration that the magnitude of the magnetic moment will determine {\TC}. 

It has been well established that information related to the crystal structure is very valuable in relation to understanding the physics of binary compounds with transition metals and rare-earth metals. Therefore, we design the third category with structural descriptors that roughly represent the structural information at the transition metal and rare-earth metal sites, which are (20) the concentration of the transition metal ($C_T$), (21) the concentration of the rare-earth metal ($C_R$), (22) the average distance between a transition-metal site and the nearest transition-metal site ($d_{T-T}$), (23) the average distance between a transition-metal site and the nearest rare-earth-metal site ($d_{T-R}$), (24) the average distance between a rare-earth metal-site and the nearest rare-earth-metal site ($d_{R-R}$), (25) the average number of nearest transition-metal sites surrounding a transition-metal site ($N_{T-T}$), (26) the average number of nearest rare-earth-metal sites surrounding a transition-metal site ($N_{T-R}$), (27) the average number of nearest rare-earth-metal sites surrounding a rare-earth-metal site ($N_{R-R}$), and (28) the average number of nearest transition-metal sites surrounding a rare-earth-metal site ($N_{R-T}$). The values of these descriptors are calculated from the crystal structures of the compounds from the literature.

\subsection{Prediction of the Curie temperature}
To learn a function for predicting the values of {\TC} of compounds from the data represented by using vectors of the descriptors, we apply the Gaussian kernel ridge regression (GKR) technique \cite{ML}. The results indicate that it is possible to accurately predict {\TC} of rare-earth transition-metal bimetal alloys by using GKR with the designed descriptors. The prediction accuracy reaches a maximum with five to ten descriptors (see Table \ref{tab.Bestdescriptors} in the Supplemental Materials). Note that for each fixed number of descriptors recruited in the GKR, a couple of different sets of descriptors also attain a high prediction accuracy almost comparable to the best one because the descriptors are not independent. By using a set of eight descriptors ($C_R$, $C_T$, $Z_R$, $Z_T$, $IP_{T}$, $S_{3d}$, $J_{3d}$, and $L_{3d}$), we can obtain an excellent prediction accuracy (as seen in Fig.~\ref{fig.Tc}) with $R^2$ and the mean absolute error (MAE) being approximately 0.96 and 41~K, respectively. 

\setcounter{table}{0}
\renewcommand{\thetable}{S\arabic{table}}

\begin{table}
\centering
\caption{Best prediction accuracy and the corresponding best descriptor sets for each given number of descriptors.}
\begin{tabular}{p{1.5cm}|p{1.9cm}|p{4.7cm}}
\hline\hline
Number of descriptors & Best prediction accuracy & Best descriptor set \\
\hline
4 & 0.950 & $C_R$, $Z_R$, $IP_T$, $S_{3d}$ \\
5 & 0.957 & $C_R$, $C_T$, $Z_R$, $L_{3d}$, $S_{3d}$  \\
6 & 0.957 & $C_R$, $C_T$, $Z_R$, $IP_T$, $L_{3d}$, $S_{3d}$  \\
7 & 0.958 & $C_R$, $C_T$, $Z_R$, $\chi_T$, $IP_T$,$L_{3d}$, $S_{3d}$,  \\
8 & 0.960 & $C_R$, $C_T$, $Z_R$, $Z_T$, $IP_T$, $S_{3d}$, $L_{3d}$, $J_{3d}$  \\
9 & 0.958 & $C_R$, $C_T$, $\chi_R$, $Z_T$, $\chi_T$, $IP_T$, $S_{3d}$, $L_{3d}$, $J_{3d}$ \\
10 & 0.959 & $C_R$, $d_{R-R}$, $Z_R$, $S_{4f}$, $Z_T$, $\chi_T$, $IP_T$, $S_{3d}$, $L_{3d}$, $J_{3d}$  \\
\hline\hline
\end{tabular}
\label{tab.Bestdescriptors} 
\end{table}

\subsection{Strong relevance and weak relevance}
On the basis of Eq.~(\ref{eq.PAS}), we can evaluate the relevance \cite{feature_selection_Yu:2004, feature_selection} of a descriptor for the prediction of {\TC} by using the expected reduction in the prediction ability caused by removing this descriptor from the full set of descriptors. Let $\bm{D}$ be a full set of descriptors, $d_i$ a descriptor, and $\bm{D}_i = \bm{D} - \{d_i$\} the full set of descriptors after removing the descriptor $d_i$. The degree of relevance of the descriptors can be formalized as follows: 

1. Strong relevance: a descriptor is strongly relevant if and only if  
\begin{equation}
PA(\bm{D})-PA(\bm{D}_i) = \max_{\forall s \subset \bm{D}} R^2_s - \max_{\forall s \subset \bm{D}_i} R^2_s > 0.
\label{eq.strong}
\end{equation}
Among the strongly relevant descriptors, a descriptor that causes a larger reduction in the prediction ability when it is removed can be considered as a strong one. The degree of relevance of a strongly relevant descriptor can be computationally estimated by using the {\bf leave-one-out} approach, i.e., by leaving out a descriptor in the currently considered descriptor set for the GKR analysis and testing how much the prediction accuracy is impaired.

2. Weak relevance: a descriptor is weakly relevant if and only if 
\begin{align}
PA(\bm{D})-PA(\bm{D}_i) = \max_{\forall s \subset \bm{D}} R^2_s - \max_{\forall s \subset \bm{D}_i} R^2_s =0 \  and \notag \\
\exists \bm{D}_i' \subset \bm{D}_i \   such \ that \  PA(\{d_i,\bm{D}_i'\})-PA(\bm{D}_i') > 0.
\label{eq.weak}
\end{align}

It is clearly seen from Eq.~(\ref{eq.weak}) that estimation of the degree of relevance for the weakly relevant descriptors cannot be carried out in a straightforward manner as for the case of the strongly relevant descriptors. Weakly relevant descriptors are descriptors that are relevant for prediction, but they can be substituted by other descriptors. We can only estimate the degree of relevance for this type of descriptor in specified contexts. For example, in terms of the prediction of {\TC}, the relevance of a descriptor for an atomic property of transition metal can be examined in the context that all of the descriptors for the atomic properties of rare-earth metals are included in the descriptor set. We define the following additional rule for comparing two weakly relevant descriptors: 

3. Comparison between weakly relevant descriptors: A weakly relevant descriptor $d_i$ is said to be more relevant than the descriptor $d_j$ in the context of having a set of descriptors $\bm{M} (d_i, d_j \notin \bm{M})$ if and only if 
\begin{equation}
PA(\{d_i,\bm{M}\}) > PA(\{d_j,\bm{M}\}).
\label{eq.compare_weak}
\end{equation}

A comparison of two weakly relevant descriptors can be computationally carried out by using the {\bf add-one-in} approach, i.e., by exclusively adding the two descriptors to the currently considered descriptor set for the GKR analysis and testing how much the prediction accuracy is improved. 

\subsection{Descriptors associated with transition-metal properties}
The evaluation of weakly relevant $\bm{T}$ descriptors measured by the improvement in the prediction ability ($\Delta R^2$) 
is summarized in Table \ref{tab.R2scores}.
The largest improvement in the prediction accuracy of {\TC} can be obtained by adding the $S_{3d}$ descriptor. The addition of the $J_{3d}$ descriptor can also yield a large improvement in the prediction accuracy of {\TC}, but the addition of the $L_{3d}$ descriptor yields a far lower improvement in the prediction accuracy. These results are consistent with the understanding so far that the values of {\TC} of binary alloys consisting of 3$d$ transition-metal (T) and 4$f$ rare-earth elements (R) are mainly determined by the magnetic interaction in the transition-metal sublattice. Indeed, in R--T compounds, there are three types of interactions including the magnetic interaction between T atoms in the T sublattices (T--T interaction), the magnetic interaction between R atoms and the T sublattices (R--T interaction), and the magnetic interaction between R atoms in the R sublattices (R--R interaction). The R--R interaction is very weak in comparison with the T--T and R--T interactions because the $4f$ electrons reside far into the interior of the rare-earth atoms and the spatial extent of the $4f$ electron wave function is rather small compared with the lattice separation. The R--T interaction is also weak in comparison to the T--T interaction; however, the R--T interaction plays an important role in determining the magnetic structure of R--T compounds. The T--T interaction dominates in R--T compounds because the delocalization and spatial extent of the $3d$ electron wave functions of T atoms are much more pronounced than those of $4f$ electrons. The understanding so far regarding the dominance of the magnetic interaction in the transition-metal sublattice is also consistent with our analysis result that the improvement in the prediction ability for the descriptors describing the magnetic properties of the constituent transition metals is obviously much larger than that for the descriptors describing the magnetic properties of the constituent rare-earth metals. 
\end{document}